\documentclass[a4paper,fleqn,useAMS,usenatbib]{mnras}

\pdfoutput=1

\usepackage{mathptmx}

\usepackage[T1]{fontenc}
\usepackage{ae,aecompl}

\usepackage[dvips]{graphicx}
\usepackage{amsmath}
\usepackage{amsfonts}
\usepackage{amssymb}
\usepackage{comment}
\usepackage[dvipsnames]{xcolor}
\usepackage[%
        ]{hyperref}
        
\hypersetup{
	colorlinks=true,	
	urlcolor=MidnightBlue,		
	pdfpagelabels=true,
	hypertexnames=true,
	plainpages=false,
	naturalnames=true,
	pdftitle={Stochastic Wow},    
	pdfauthor={Kipping and Gray},     
	linkcolor=WildStrawberry,          
	citecolor=ForestGreen,        
}

\newcommand{\pdf}{\mathrm{Pr}}

\newcommand{\wwwcoolworlds}{\href{https://github.com/davidkipping/wow/}{this URL}}

\newcommand{\Tmin}{T_{\mathrm{min}}}
\newcommand{\Tmax}{T_{\mathrm{max}}}
\newcommand{\Lmin}{\lambda_{\mathrm{min}}}
\newcommand{\Lmax}{\lambda_{\mathrm{max}}}

\title[``Wow'' as a stochastic repeater]{
Could the ``Wow'' signal have originated from a stochastic repeating beacon?
}
\author[Kipping \& Gray]{David Kipping$^{1}$\thanks{E-mail:
\href{mailto:dkipping@astro.columbia.edu}{dkipping@astro.columbia.edu}} and Robert Gray$^{2}$\\
$^{1}$Dept. of Astronomy, Columbia University, 550 W 120th Street, New York NY 10027\\
$^{2}$Gray Data Consulting, 3071 Palmer Square, Chicago, IL 60647}

\date{Accepted . Received ; in original form }

\pubyear{2022}

\begin{document}
\label{firstpage}
\pagerange{\pageref{firstpage}--\pageref{lastpage}}
\maketitle

\begin{abstract}
The famous ``Wow'' signal detected in 1977 remains arguably the most compelling
SETI signal ever found. The original Big Ear data requires that the signal
turned on/off over the span of ${\sim}3$\,minutes (time difference between the
dual antennae), yet persisted for 72\,seconds (duration of a single beam
sweep). Combined with the substantial and negative follow-up efforts, these
observations limit the allowed range of signal repeat schedules, to the extent
that one might question the credibility of the signal itself. Previous work has
largely excluded the hypothesis of a strictly periodic repeating source, for
periods shorter than 40\,hours. However, a non-periodic, stochastic repeater
remains largely unexplored. Here, we employ a likelihood emulator using the Big
Ear observing logs to infer the probable signal properties under this
hypothesis. We find that the maximum \textit{a-posteriori} solution has a
likelihood of 32.3\%, highly compatible with the Big Ear data, with a broad
2\,$\sigma$ credible interval of signal duration $72$\,secs$<T<77$\,mins and
mean repeat rate $0.043$\,days$^{-1}<\lambda<59.8$\,days$^{-1}$. We extend our
analysis to include 192\,hours of subsequent observations from META, Hobart and
ATA, which drops the peak likelihood to 1.78\%, and thus in tension with the
available data at the 2.4\,$\sigma$ level. Accordingly, the Wow signal cannot
be excluded as a stochastic repeater with available data, and we estimate that
62\,days of  accumulated additional observations would be necessary to surpass
3\,$\sigma$ confidence.
\end{abstract}


\section{Introduction}

In 1977, the Ohio State University Radio Observatory recorded a strong
(30\,$\sigma$), narrowband ($<10$\,kHz) signal near the 21\,cm hydrogen line
\citep{kraus:1979}. The signal, discovered during a search for extraterrestrial
intelligence (SETI), bore a remarkable resemble to the properties expected
from an alien transmitter, leading the volunteer analyst Jerry Ehman to
famously annotate ``Wow!'' on the computer printout \citep{ehman:1998}.
The time-dependent intensity of the emission matched the antenna pattern
signature of a transiting celestial source \citep{gray:2001}, unlike what
would be expected for interference. This behaviour, combined with its
narrowband nature and location near 21\,cm, leads to a trifecta of features
long predicted from a genuine SETI signal \citep{coccini:1959}, causing Wow to
be one of the few astronomical observations that have bled into popular
culture.

Although the field was observed many times, Wow was only seen once during the
entire baseline of observations from the Ohio State University Radio
Observatory, known as the ``Big Ear'' \citep{kraus:1979}. Further, Wow only
occurred in one beam of a dual-beam transit antenna system \citep{dixon:1977},
implying that the signal switched on/off in the brief interval between the two
beams sweeping over the source ($\sim 3$\,minutes). However, Wow appears
consistent with being continuously ``on'' during the 72\,seconds of one beam
passing over the source. 

Despite numerous attempts to replicate the detection of Wow \citep{gray:1994,
gray:2001,gray:2002,harp:2020}, the signal has not been observed again. The
single-beam detection and lack of any repetition (especially with greater
sensitivity, longer observations and broader spectral coverage), ostensibly
places considerable pressure on the credibility of the Wow signal\footnote{For
example, in an interview with the
\href{Cleveland Plain Dealer}{http://www.bigear.org/wow.htm} in 1994, Jerry
Ehman remarked ``We should have seen it again when we looked for it 50 times.
Something suggests it was an Earth-sourced signal that simply got reflected off
a piece of space debris.''}. However, a recent Bayesian formulation of one-off
events showed that at least an order-of-magnitude more data than that used in
the original discovery is often required to place statistically significant
tension on a signal's credibility \citep{blackswans}.

Motivated by that work, we here explore a rigorous statistical
approach to address: i) how likely is it that Wow repeats but has simply been
missed by follow-up efforts to date, and ii) if so, what are the most likely
properties of the signal under this scenario?

There are two basic ways that Wow could remain a genuine SETI signal yet remain
consistent with all observations to date. It could either be a non-continuous
emission source and/or a continuous source that drifts in frequency. Regarding
the latter possibility, Wow was detected in channel 2 of 50 by Big Ear
\citep{kraus:1979}, each with a 10\,kHz band. Given that the signal
appears/disappears over a 3\,minute interval, a drift rate of
20\,kHz/3\,minutes $\simeq 100$\,Hz\,s (i.e. $\sim 100$\,nHz) could conceivably
explain this\footnote{Such a signal would drift 7.2\,kHz during the 72\,second
duration of a single-beam, and thus would also plausibly be confined to a
single 10\,kHz channel.}. We note that this would be considered at the upper
fringes of plausibility for a genuine SETI signal \citep{sheikh:2019}; for
example the Earth's fractional drift rate from orbital and rotational motion is
three orders-of-magnitude smaller. Further, subsequent observations by
\citet{gray:2002} used a band five times wider than than of Big Ear (500\,kHz)
spanning six 14-hour continuous blocks, yet reported no Wow-like repetition.

A non-continuous signal is more interesting and might be expected from an
economics perspective, as suggested by \citet{benford:2008}. A strictly
periodic repeater might be expected if the source behaved akin to a lighthouse,
perhaps as a continuous source rotating across the sky or programmed to
emit only periodically in our direction. \citet{gray:2002} and
\citet{harp:2020} excluded strictly periodic repeaters with periods below
40\,hours and longer timescales become increasingly improbable to fortuitously
allow the Wow signal being detected in the first place. To date, there has been
no investigation of a non-periodic, \textit{stochastic} repeater though. A
stochastic repeater could be attractive from the perspective of avoiding
sampling holes. For example, signals that repeat once per 24\,hours both risk
evading our observing cadence as well being immediately suspicious as spurious
even if recorded. An intermittent signal could also manifest due to scintillation within the interstellar medium, in particular for sources $>100$\,parsecs away \citep{cordes:1997}. The underlying cause of the intermittency is not of principal interest here, rather we consider that such intermittency is feasible and the compatibility of existing observations with such a scenario has not been extensively studied for Wow.

In this work, we investigate this possibility by evaluating the
plausibility of said hypothesis and the corresponding required signal
properties. Although \citet{blackswans} presents a framework for such an
analysis, that work is only valid when the data are homogenous, regular and no
substantial changes to the observing instruments nor strategy are implemented.
In order to account for the sparse, irregular sampling of Big Ear, as well
as the subsequent observations, we require a more detailed approach that
emulates the detection efficiency of the instrument for various proposed
signal forms. To this end, Section~\ref{sec:likelihood} presents such an
emulator, which is then used in Section~\ref{sec:posterior} to infer a
posterior distribution for the signal form. We extend this analysis
in Section~\ref{sec:followup} to include observations besides from that of
Big Ear, and conclude in Section~\ref{sec:discussion}.

\section{The likelihood of a Wow-like signal}
\label{sec:likelihood}

\subsection{Model for the Signal(s)}

Searches for strict periodicity can be readily achieved using classic
periodogram techniques. Stochastic repeaters, in contrast, may evade
detection using such techniques. Accordingly, we here seek to test
the specific hypothesis of a stochastic repeater.

In what follows, we assume that the source maintains a uniform rate of signal
production over any fixed interval of time within the baseline of the Big
Ear observations. This essentially corresponds to a kind of Copernican
Principle in time - we assume that the Big Ear observations do not occupy a
special point in time when the signal source changed its emission mode (for
example). Indeed, this assumption demands that the mean number of signals
produced per unit time (in our direction) is the same when Big Ear began
observing the source, as when it finished. This defines a Poisson process, for
which the interval between individual signals will follow an exponential
distribution. Accordingly, the $(k+1)^{\mathrm{th}}$ signal will occur at a
time

\begin{align}
t_{\mathrm{signal},k+1} \sim t_{\mathrm{signal},k} + \mathcal{E}[\lambda],
\label{eqn:times}
\end{align}

where $\mathcal{E}[\lambda]$ is an exponential distribution of mean
$\lambda$. For a given choice of $\lambda$, we thus begin our simulation
by iteratively populating a list of signal mid-times using 
Equation~\ref{eqn:times}, where we start with $t_{\mathrm{signal},0}=0$ but
ultimately discard this initial entry.

The signal is also characterised by a duration. In what follows, we assume
that the signal is a top-hat persisting for a time $T$ (with no variability),
since this is both consistent with the Wow signal and represents the simplest
possible model. If $t_{\mathrm{signal},k+1}-t_{\mathrm{signal},k} < T$, the
two signals are effectively merged into a longer continuous one. 

\subsection{Model for the Observations}

The observations are modelled using the observing logs of the Wow field
from Big Ear itself. In September 1985, co-author Robert Gray visited
OSU where the logs were reviewed and noted. Print outs were pulled from the
OSU archives by Marc Abel, who located runs from 122 days spanning August
1977 to March 1984. Data near the Wow locale was examined for each day, with
the locale defined as near RA 19 17 in 1977, and RA 19 22 and 19 25
subsequently, since a horn squint correction had presumably been applied by
1978.

Excluding observations more than 20' north or south of the Wow declination, or
days with bad data (e.g. clock off, printer failure, power failure), we find 90
useful visits, which are listed in Table~\ref{tab:tab1}.

These dates are used to define a set of representative observations in our
simulation, with the first observation occurring at
$t_{\mathrm{obs},1} = 0.5$\,days, and subsequent observations offset in time
from this reference point according to their observation date. Each observation
in fact a pair of measurements, since the Big Ear horns are physically separated
and thus a sky source passes over each. Accordingly, we define the first of these
two occurring at a mid-time of $t_{\mathrm{obs},l}$ (for the $l^{\mathrm{th}}$
observation) and the next occurring at $t_{\mathrm{obs},l} + \Delta t$, where
$\Delta t = 172.37$\,seconds represents the temporal offset.

\begin{figure*}
\begin{center}
\includegraphics[width=17.0cm,angle=0,clip=true]{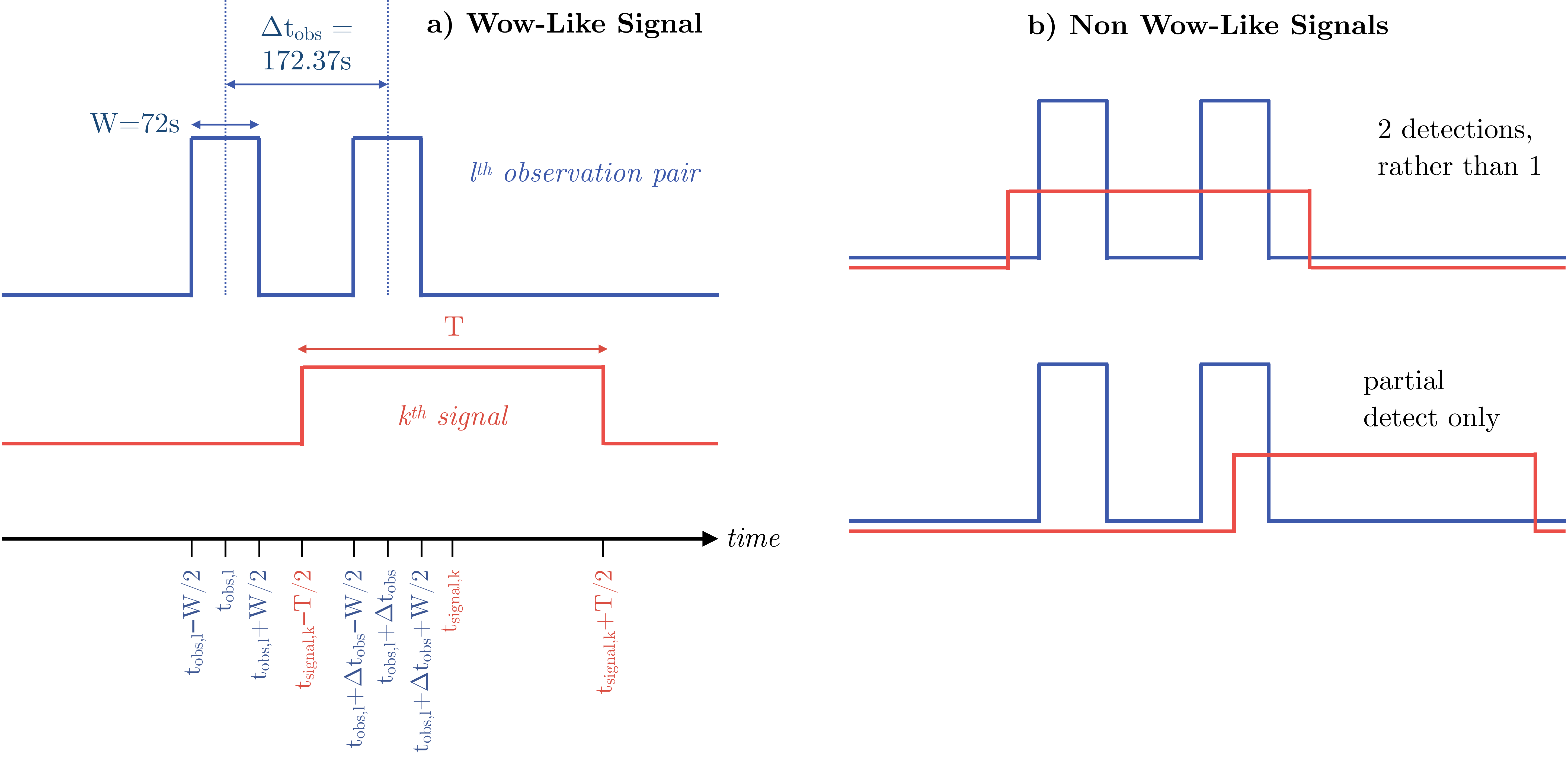}
\caption{
a) An annotated example of a Wow-like signal detection. The observational
window is shown in blue, and the signal in red. Here, the signal is fully
detected but only in one of the windows, not the other, just like Wow.
b) Examples of non-detections, either because of too many (top) or
too few (bottom) signal overlaps with the observing windows.
}
\label{fig:cartoon}
\end{center}
\end{figure*}

Each of the pair of observations is modelled as lasting for 72 seconds. For
a ``detection'' to occur, we define that a signal must be throughout the entire
duration of this 72-second window. We thus cycle through each observation
($90 \times 2$ in total) and query whether any of the generated signals
lead to a detection. After cycling through all observations, we tally up the
total number of detections. Illustrative examples of both Wow-like and non
Wow-like signals are depicted in Figure~\ref{fig:cartoon}.

\subsection{The Likelihood of a Wow-Like Signal}

The process of generating a set of injected signals, comparing them to our
simulated observations and tallying up the number of detections defines a
single realisation. Since the signal generation process is stochastic, a
statistical interpretation requires repeating many realisations for a given
choice of $T$ and $\lambda$. In practice, we approach this by creating a
``while'' loop that generates many realisations and continues until
exactly one detection is made - since this is what was found for the Wow
signal. The greater the number of realisations necessary, the less likely
it is that the particular choice of $T$ and $\lambda$ could reproduce Wow.
Naively, one might reason that if it took $C$ realisations to get a Wow-like
signal, then the probability of $T$ and $\lambda$ producing Wow is $1/C$.

For computational efficiency, we cap the maximum number of realisations in
our while loop at 1000 and flag such cases appropriately. Although this
single while loop provides a first estimate of $1/C$ probability, it is
inherently uncertain given the fact only a single loop was executed. Thus,
we wish to repeat the loop $W$ times, in order to improve our confidence
in the estimated probability.

Since our while loop represents a binary outcome of fixed
probability, then $C$ represents the number of Bernoulli trials needed to
get one success, defining a geometric distribution i.e. $C \sim 
\mathcal{G}[p]$, where $\mathcal{G}$ represents the geometric
distribution and $p$ is the probability value for this choice of
$T$ and $\lambda$. We now wish to infer $p$ from a sequence
of $C$ evaluations emerging from our while loops, which form a vector
with indices $C_w$ of length $W$. To accomplish this, we note that

\begin{align}
\pdf(\{C_1,C_2,...,C_W\}|p) &= (1-p)^{C_{\mathrm{tot}}} p^W,
\end{align}

where $C_{\mathrm{tot}} \equiv \sum_{w=1}^W C_w$. To infer
$\pdf(p|\{C_1,C_2,...,C_W\})$ (the posterior distribution of $p$),
we need to use Bayes theorem. Adopting a uniform prior on $p$
for simplicity, the posterior is simply the normalised likelihood
given by

\begin{align}
\pdf(p|\{C_1,C_2,...,C_W\}) &= \frac{\Gamma[2+W+C_{\mathrm{tot}}](1-p)^{C_{\mathrm{tot}}} p^W}{\Gamma[W+1] \Gamma[C_{\mathrm{tot}}+1]}.
\end{align}

Differentiating, this has a mode (i.e. maximum \textit{a-posteriori}, MAP)
value of

\begin{align}
\bar{p} &= \frac{W}{W+C_{\mathrm{tot}}},
\end{align}

which is close to our original intuition of $1/C_{\mathrm{tot}}$ for $W\sim1$.
However, this process also allows us to define the variance to be

\begin{align}
\sigma_p^2 &= \frac{(W+1)(C_{\mathrm{tot}}+1)}{(2+W+C_{\mathrm{tot}})^2(3+W+C_{\mathrm{tot}})}.
\end{align}

Using these results we can say that after $W$ while loops, each of which took
$C_w$ realisations to obtain a Wow-like signal, we infer the MAP likelihood of
the choice of $T$ and $\lambda$ to be $\bar{p}$ with a precision (fractional
error) of $\bar{p}/\sigma_p$. We thus continue through the while loops until
the precision falls below 10\% - which we qualify as a ``good'' estimate of
$\bar{p}$.

\subsection{Likelihood Emulator}

From before, $\bar{p}$ represents the maximum \textit{a-posteriori} probability
of obtaining a Wow-like signal with the 90 days of Big Ear data, given a choice
of $\lambda$ and $T$. In moving towards inferring $\lambda$ and $T$,
$\bar{p}$ can be considered to be the (most probable) likelihood of obtaining
the data $\mathcal{D}$ (here defined as a single detection in the 90 pairs
of observations) given $\lambda$ and $T$. Thus, we define

\begin{align}
\pdf(\mathcal{D}|\lambda,T) &= \frac{W}{W+C_{\mathrm{tot}}}.
\end{align}

To make progress, we need to be able to vary $\lambda$ and $T$ and
evaluate likelihoods quickly, to exploit Bayesian inference methods. Given
the computational expensive nature of even a single likelihood evaluation
(sometimes taking many hours), we elected to build a grid of $\lambda$ and $T$
values across which we could pre-calculate the likelihoods and then use
them to build a likelihood emulator at intermediate values. To this end,
we set up a $M=100$ square grid of plausible values for each. The grid was
chosen to be log-uniformly spaced from $\Lmin$ to $\Lmax$ and $\Tmin$ to
$\Tmax$, such that

\begin{align}
T_i= \exp\Big[ \log\Tmin + (i-1) \frac{\log\Tmax-\log\Tmin)}{M-1} \Big],\nonumber\\
L_j= \exp\Big[ \log\Lmin + (j-1) \frac{\log\Lmax-\log\Lmin)}{M-1} \Big],
\end{align}

where $i$ and $j$ are the integer indices running from 1 to $M$.
We set $\Tmin = 72$\,seconds, which defines the minimum allowed duration given
that the signal shape appears consistent with a continuous emission. For
$\Tmax$, we chose a maximum at least two orders of magnitude larger and
specifically adopted $\Tmax=e^{-1}$\,days. For $\Lmin$ and $\Lmax$, we chose to
bracket them three orders of magnitude either side of $1/\Tmax$, giving
$\Lmin = e^{-4}$\,days$^{-1}$ and $\Lmax=e^2$\,days$^{-1}$.

Our later parameter explorations found that this initial grid did not support
certain regions of high likelihood space and needed to be expanded. To
accomplish this, we simply increased the indices $i$ and $j$ beyond 100.
Specifically, we ran up to $i=150$ and $j=160$. This means that in practice the
maximum allowed $T$ value is not in fact $\Tmax$, but rather
$\Tmax'$, and similarly for $\lambda$, where

\begin{align}
\Tmax' &= \exp\Big[ \log\Tmin + \frac{149}{99} (\log\Tmax-\log\Tmin),\nonumber\\
\Lmax' &= \exp\Big[ \log\Tmin + \frac{159}{99} (\log\Tmax-\log\Tmin),
\end{align}

which evaluates to $\Tmax' = 7.97$\,days and $\Lmax' = 280.4$\,days$^{-1}$.

\begin{figure*}
\begin{center}
\includegraphics[width=17.0cm,angle=0,clip=true]{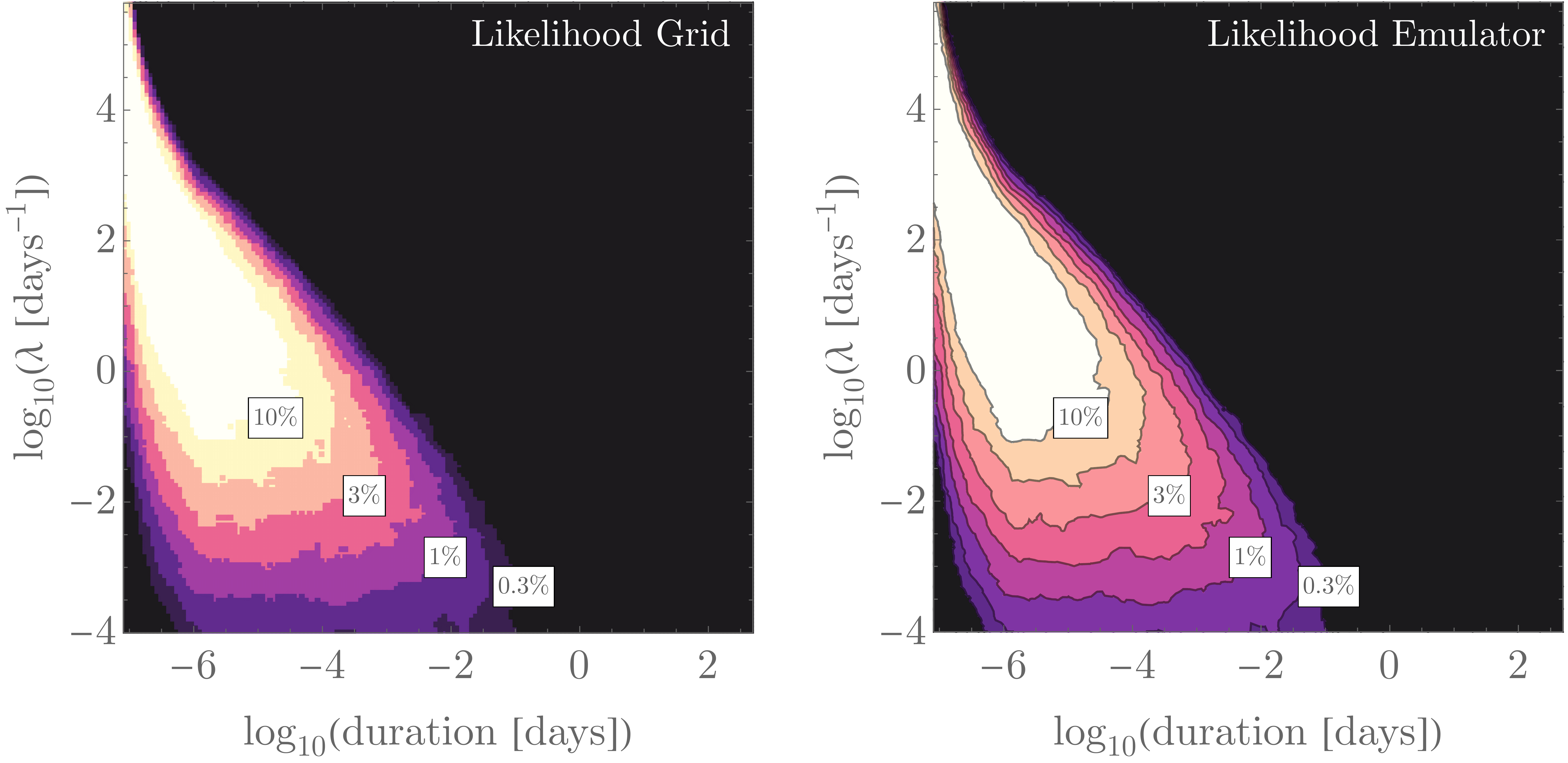}
\caption{
Comparison of our grid of individual likelihood evaluation (left) and the
emulator used for the MCMC using interpolation (right). We colour code
the points/regions by their likelihood value, with the highest region
exceeding 10\%.
}
\label{fig:likegrid}
\end{center}
\end{figure*}

Since our grid is regular and evenly spaced (in log space), it is amenable
to interpolation. To this end, we use a bi-cubic spline interpolation of the
grid. As shown in Figure~\ref{fig:likegrid}, this leads to an accurate and
quick emulator across the high likelihood region of interest.

\section{The \textit{a-posteriori} ``Wow'' signal properties}
\label{sec:posterior}

\subsection{MCMC}

Having established a likelihood emulator supported over the range
$\Tmin\leq T < \Tmax'$ and $\Lmin\leq \lambda < \Lmax'$, we now
turn to inferring the joint posterior distribution of $T$
and $\lambda$, given the data $\mathcal{D}$. This can be achieved
through Bayes theorem:

\begin{align}
\pdf(T,\lambda|\mathcal{D}) \propto \pdf(\mathcal{D}|T,\lambda) \pdf(T,\lambda).
\end{align}

For the prior, $\pdf(T,\lambda)$, we assume that the priors are independent
and follow log-uniform distributions over their supported range. This is most
easily achieved by simply walking in $\log T$ and $\log \lambda$ space
directly, which is also convenient as this is the native sampling the
grid used for the likelihood emulator. The limits on the walkers are set to
that of our likelihood grid, since outside this region we cannot reliably
emulate the likelihood.

We sample the joint posterior distribution using Markov Chain Monte Carlo
algorithm and Metropolis walkers, calculating the posterior probability
as a product of our likelihood emulator and the prior (which is uniform
in this parameterisation). Each walker completes 110,000 accepted
steps, burning off the first 10,000. The starting point for the chain
is randomly drawn from the joint prior. We repeat for 100 walkers and
combine the final results, with a thinning of a factor of 10 to
leave us with $10^6$ posterior samples. Chains and posterior distributions
were inspected to ensure convergence and good mixing.

\subsection{Analysis}

\begin{figure*}
\begin{center}
\includegraphics[width=17.0cm,angle=0,clip=true]{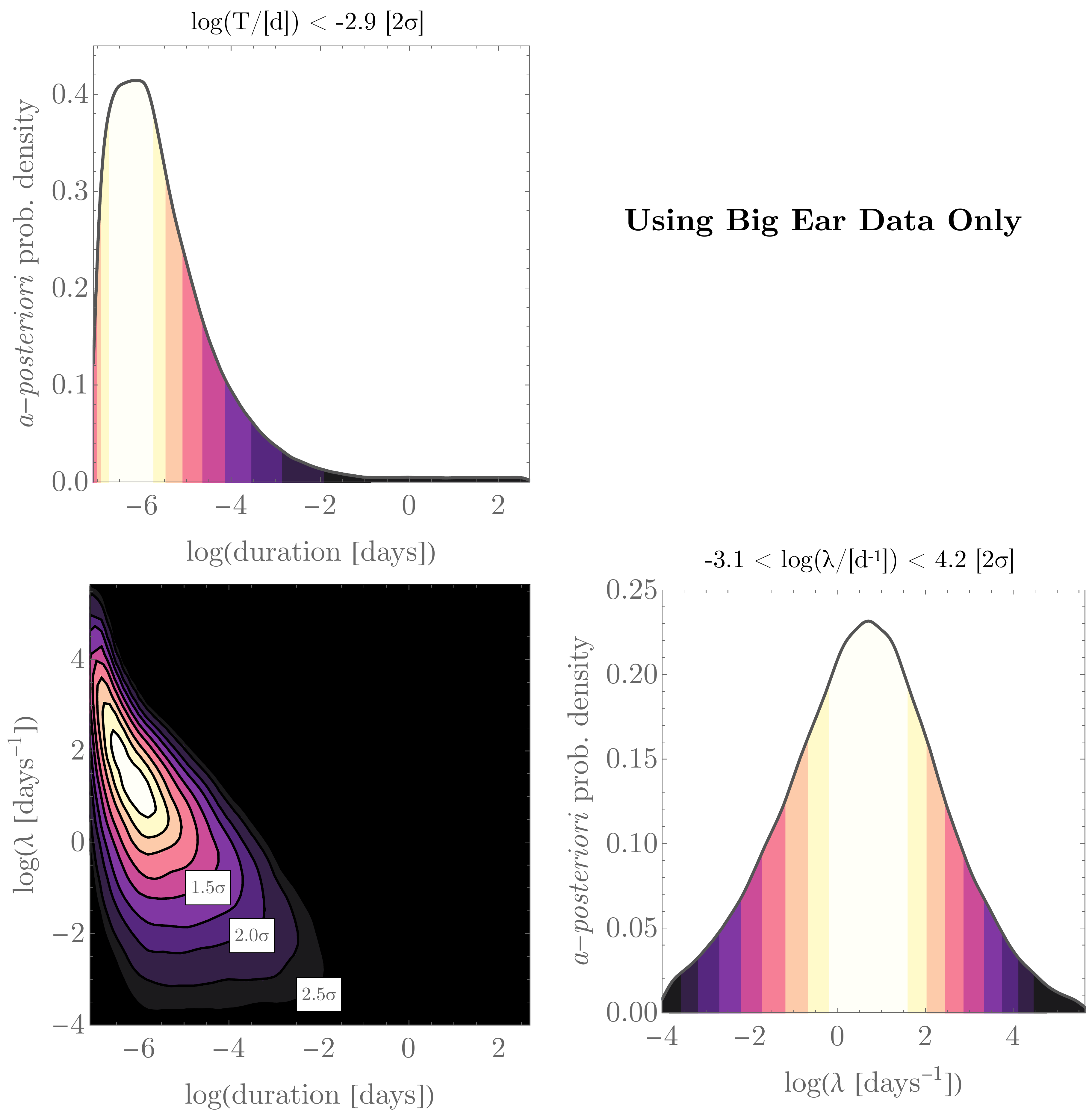}
\caption{
Triangle plot of the posterior distribution for the Wow signal properties.
The joint posterior (bottom-left) naturally has a similar shape to the
likelihood (Figure~\ref{fig:likegrid}) given our diffuse prior, but the
contours now represent the \textit{a-posteriori} credible intervals rather
than likelihood confidence limits. Contours are separated by 0.25\,$\sigma$.
}
\label{fig:post1}
\end{center}
\end{figure*}

The joint posterior distribution is depicted in Figure~\ref{fig:post1}.
We find that the marginalised duration distribution peaks close to the
lower boundary condition. Since MCMC samplers cannot make jumps below the
prior minimum threshold, a slightly positively skewed posterior is
expected even when the result is consistent with the minimum. Accordingly,
we here conclude that there is no clear lower limit on the signal duration,
besides from the constraint that it appears consistent with being fully-on
during the window in which it was caught (i.e. $T>72$\,seconds). However,
we find that long duration signals are disfavoured, which can be understood
by the fact that they are increasingly unlikely to have only been detected
once during the ensemble of Big Ear observations. Indeed, from this, we
can place an upper limit that $T<77$\,minutes to $2$\,$\sigma$ confidence
(95.45\%).

For $\lambda$, the mean rate of signal emission, we find a broad but peaked
posterior with a maximum at $\lambda = 2.01$\,days$^{-1}$ (i.e. twice per day).
We can see that $\lambda$ is inversely correlated to $T$ but compresses
in a narrow region at low $T$. This ``fine-tuning'' reduces the marginalised
posterior density in this region, and ultimately gets truncated by the
lower boundary on $T$. Very frequent repeat times are only compatible with
the data at extremely short durations, in order to explain the lack of
additional detections. Together, these constraints imply that $\lambda$
is bound within the interval $0.047$\,days$^{-1}$ to $64.2$\,days$^{-1}$
to 2\,$\sigma$.

\section{Accounting for additional follow-up}
\label{sec:followup}

\subsection{Follow-up Observations}

Of course, besides from the original Big Ear observations, there have been
several efforts to re-observe the field and look for repetitions, without
success. In \citet{gray:1994}, the field was observed using the ultranarrowband
Harvard/Smithsonian-META system \citep{horowitz:1986}, tracking for up to
4\,hour at a time. \citet{gray:2001} performed the second attempt with
the Very Large Array, tracking the field for no more than 22\,minutes a
time. \citet{gray:2002} performed far longer observations with 
the University of Tasmania's Hobart 26\,m radio telescope, tracking the
field for 14\,hours continuously on six separate days. Finally, the
Allan Telescope Array was used by \citet{harp:2020}, accumulating over
100\,hours on target for 30 to 180\,minutes each time. In all cases,
we note that the observations were sufficiently sensitive to have
recovered a repeat of the Wow signal to high confidence.

\subsection{Incorporating the Hobart Constraints}

Amongst the four follow-up efforts highlighted above, the \citet{gray:2002}
data with Hobart is most amenable to incorporating into our analysis,
since the observing window is approximately the same each time (14 hours)
and only six sets were taken. The dates of the observations are provided
in \citet{gray:2002} and thus we could easily imagine directly extending
our likelihood formalism to include these dates. However, a much simpler
approach would be to ignore the dates and rather just assume that each
14-hour window is an independent and fair sample over all time within
which no detection was made. For a $F=14$\,hour block, the probability
of a Poisson process not yielding a single success equals
$e^{-F \lambda}$, and to evade detection in six independent draws would
be $(e^{-F \lambda})^6 = e^{-6 F \lambda}$. This serves as an additional
likelihood term which we can append to the Big Ear likelihood emulator.
With this simple change, we repeated the MCMC analysis using an otherwise
identical setup.

The updated chain found that the maximum likelihood dropped from 32.3\%
using Big Ear alone, to 3.27\%, thus placing tension on the stochastic
repeater hypothesis at the 2.1\,$\sigma$ level. This peak occurs
at $\hat{\lambda} = 0.233$\,days$^{-1}$ and $\hat{T} = 485$\,s. Evaluating
the likelihood of this position using the Big Ear likelihood emulator
alone (ignoring the Hobart data) yields 7.40\%. Thus, the likelihood
shifts from 7.40\% to 3.27\% - a factor of 0.442 of the original value,
which of course simply equals $e^{-6 F \hat{\lambda}}$. Using this point
of highest likelihood, we can test the accuracy of our assumption of
independence.

To this end, we used the dates of the Hobart observations and generated
1000 Poisson processes with a mean rate $\hat{\lambda}$ and duration
$\hat{T}$ and counted how many detections Hobart would have seen.
From this, we measure 439 zero-count cases - consistent with the real
observations of \citet{gray:2002}. Thus, we find that a more realistic
calculation that accounts for the specific timings of the observations
yields a Hobart-likelihood at $\hat{\lambda}$ and $\hat{T}$ of
$(0.439\pm0.016)$ - a value fully consistent with the much simpler
independence assumption. On this basis, we argue that our simplifying
assumption is justified and produces an accurate final posterior.

From our revised joint posterior, we obtain credible intervals of
$\lambda = [0.021,0.441]$\,d$^{-1}$ and
$T = [130\,\mathrm{s},2.2\,\mathrm{d}]$.

\subsection{Extending to Other Observations}

Our approach for incorporating the Hobart observations can be extended
to other data sets too. Recall that for Hobart, the additional likelihood
term was $(e^{-F \lambda})^6 = e^{-6 F \lambda}$, since 6 sets of $F$
baselines were taken. This is equivalent to adding a single penalty of
$6F$ duration (i.e. 84\,hours). Accordingly, even if the other observations
did not use the same observational window each, we can still include
them by simply counting up the total observing time of each (on the Wow
field).

From \citet{gray:1994}, we find 8\,hours on each of the two possible
Wow positions. \citet{gray:2001} dwelled no longer than 22\,minutes on
the field and used a variety of observing modes, and thus we elected to
not include these data as their temporal constraining power is small
compared to Hobart. In contrast, the ATA 100\,hour campaign from
\citet{harp:2020} more than doubles the baseline and thus provides a
key constraint. Together then, we add a third likelihood term equal
to $e^{-G \lambda}$, where $G = 108$\,hours.

\begin{figure*}
\begin{center}
\includegraphics[width=17.0cm,angle=0,clip=true]{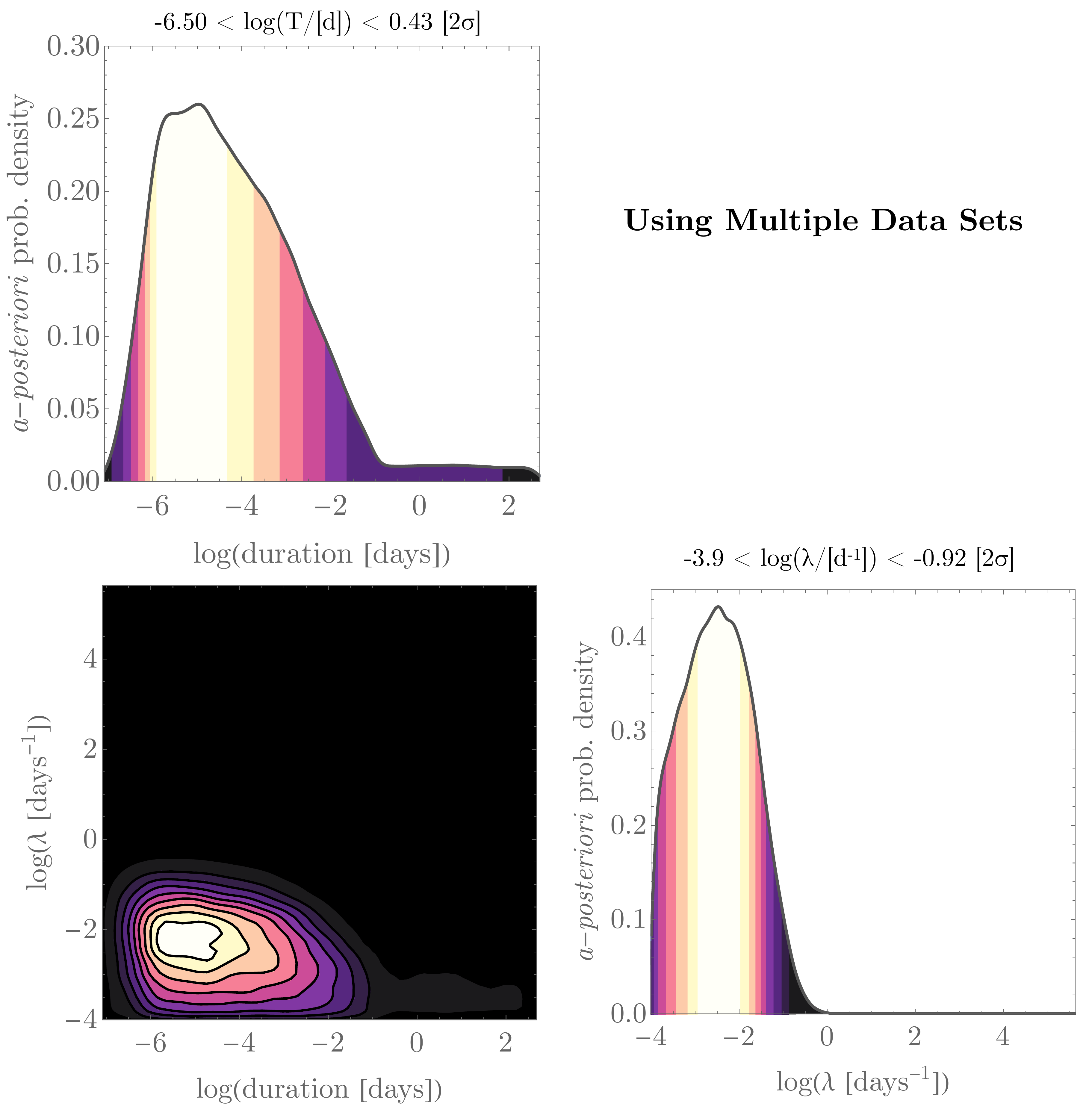}
\caption{
Same as Figure~\ref{fig:post1}, but the MCMC now includes additional
constraints from the Harvard/Smithsonian-META system \citep{gray:1994},
the University of Tasmania's Hobart 26\,m radio telescope \citep{gray:2002}
and the Allan Telescope Array \citep{harp:2020}.
}
\label{fig:post3}
\end{center}
\end{figure*}

With this change, the final maximum likelihood is 1.78\% (2.37\,$\sigma$),
occurring at $\hat{T} = 659$\,s and $\hat{\lambda} = 0.121$\,days$^{-1}$.
The final joint posterior is plotted in Figure~\ref{fig:post3}, where we report
a revised 2\,$\sigma$ credible interval of
$T = [130\,\mathrm{s},1.53\,\mathrm{d}]$ and
$\lambda = [0.021,0.397]$\,d$^{-1}$. In practice, the joint posterior is
not significantly changed by the inclusion of the other observational
constraints.

\section{Discussion}
\label{sec:discussion}

Our work has used a likelihood emulator derived from the Big Ear observing logs to
investigate the properties of the Wow signal, under the assumption it is a
stochastic repeater. Here, we discuss the implications of this result, and
validity of our underlying assumption.

We'll begin with the latter. It must be acknowledged that have no evidence that
the Wow signal is a repeating signal. Indeed, the lack of any repeat is what makes
the signal so tantalising and ultimately motivates our study to investigate how
viable the repeating hypothesis is. However, even if the signal is non-repeating,
it is technically consistent with our formalism in the limiting case that
the mean repeat time tends to infinity (i.e. $\lambda \to 0$). With just a single
pulse produced over all cosmic time, it's incredibly improbable that the Big Ear
would have coincidentally been looking at the correct place at the correct moment
to catch it. Indeed, our analysis with the Big Ear data alone has a marginalised
$\lambda$ posterior density that tends to 0 in the limit of $\lambda \to 0$.

Using just the Big Ear data, strict periodic signals are challenging to
constrain since the field was observed briefly just twice per day, meaning that
periodic signals can easily fall out of phase with an initial detection. This
possibility was investigated in \citet{gray:2002}, although the duration of the
signal was fixed to three scenarios of $T=72$\,s, $T=144$\,s and $T=288$\,s,
rather than freely fitted as done in this work. \citet{gray:2002} report that
periods of below one day had a $>5$\% likelihood of reproducing the
observations, peaking at 35\% for signals of approximately an hour period.

To compare to our stochastic repeater analysis, we find a peak likelihood of
32.3\% for a signal of duration $\hat{T}=181$\,s and mean rate
$\hat{\lambda}=3.69$\,days$^{-1}$ ($\sim 6.5$\,hours mean repeat time). Thus,
the stochastic assumption can be seen to allow for even higher compatibility
with the original data, although the timescales involved of the best fitting
signals are not grossly different. This of course is not surprising given the
inherently more ``slippery'' nature of a stochastic process.

Of course, follow-up observations have placed much stronger constraints on
strictly periodic signals. \citet{gray:2002} completely excludes strict
periodicity for periods less than 14\,hours. Going further, \citet{harp:2020}
provides the tightest constraints to date, finding that periods from 0 to
40\,hour would have been detected in 99\% of their realisations.

As shown in Section~\ref{sec:followup}, including these additional observations
into our analysis provides tighter constraints on the joint posterior. Most
importantly, the inclusion of this data reduces the maximum likelihood
from 32.3\% with Big Ear alone to 1.78\% (2.37\,$\sigma$). Thus, the data
collected to date are surprising under the stated hypothesis, but not at the
level typically needed to reject a hypothesis ($>3$\,$\sigma$ or more). We
repeated the follow-up MCMC but adding $X$\,hours of simulated null results to
investigate how much data would be needed to reach a 3\,$\sigma$ threshold.
Iterating, we found that 62\,days of accumulated additional observations would
be needed\footnote{Note that this timescale is insensitive to the priors used since it represents simply the maximum likelihood value.}. On this basis, we argue that the stochastic repeater hypothesis is
not yet excluded as an explanation for the Wow signal, and yet more follow-up
would be necessary.

More broadly, our likelihood emulation approach provides an example of
dealing with sparse, irregular observations to conduct inference of SETI
signals of interest. Indeed this approach could be readily applied to other technosignature data sets, such as the Breakthrough Listen radio search \citep{price:2020,sheikh:2021} or even non-radio searches such as in the optical (e.g. see \citealt{maire:2020}). It is emphasised that such an approach could also be used for targets without
any signals detected, in order to derive robust upper limits on a hypothesised
signal's properties (be it periodic or not).

\section*{Acknowledgments}

Co-author Robert Gray tragically passed away during this work. RG devoted
himself to pursuing the Wow signal throughout his career, both in science
and public discourse. His insights and contributions to this enigma are
unparalleled and were pivotal to this paper. DK is grateful for his generosity,
time and contributions to this study, which it is hoped live up to his
high standards of inquiry and curiosity.

This work was enabled thanks to supporters of the Cool Worlds Lab, including
Mark Sloan,
Douglas Daughaday,
Andrew Jones,
Elena West,
Tristan Zajonc,
Chuck Wolfred,
Lasse Skov,
Graeme Benson,
Alex de Vaal
Mark Elliott,
Methven Forbes,
Stephen Lee,
Zachary Danielson,
Chad Souter,
Marcus Gillette,
Tina Jeffcoat,
Jason Rockett,
Scott Hannum,
Tom Donkin,
Andrew Schoen,
Jacob Black,
Reza Ramezankhani,
Steven Marks,
Philip Masterson,
Gary Canterbury,
Nicholas Gebben,
Joseph Alexander,
Mike Hedlund,
Dhruv Bansal,
Jonathan Sturm,
Rand Corporation,
Ian Attard \&
Leigh Deacon.

\section*{Data Availability}

The code used and results generated by this work are made publcily available
at \wwwcoolworlds.

\appendix

\begin{table*}
\caption{
Summary of OSU reobservations of the Wow locale, 1977-1984. Data
comes from printouts pulled by Marc Abel around September 1985,
and then filtered against bad data (e.g. clock errors, printer
failures, etc) by co-author Robert Gray. Wow detection date
is highlighted in bold.
} 
\centering 
\begin{tabular}{c c c c c} 
\hline\hline 
Date & Days Passed & \vline & Date & Days Passed \\
\hline
13-Aug-1977 & 0 & \vline & 22-Jan-1983 & 1988 \\
14-Aug-1977 & 1 & \vline & 22-Jan-1983 & 1988 \\
\textbf{15-Aug-1977} & 2 & \vline & 23-Jan-1983 & 1989 \\
17-Aug-1977 & 4 & \vline & 24-Jan-1983 & 1990 \\
16-Sep-1977 & 34 & \vline & 25-Jan-1983 & 1991 \\
17-Sep-1977 & 35 & \vline & 28-Jan-1983 & 1994 \\
18-Sep-1977 & 36 & \vline & 29-Jan-1983 & 1995 \\
19-Sep-1977 & 37 & \vline & 2-Feb-1983 & 1999 \\
20-Sep-1977 & 38 & \vline & 2-Feb-1983 & 1999 \\
21-Sep-1977 & 39 & \vline & 4-Feb-1983 & 2001 \\
22-Sep-1977 & 40 & \vline & 5-Feb-1983 & 2002 \\
23-Sep-1977 & 41 & \vline & 6-Feb-1983 & 2003 \\
24-Sep-1977 & 42 & \vline & 7-Feb-1983 & 2004 \\
25-Sep-1977 & 43 & \vline & 12-Feb-1983 & 2009 \\
26-Sep-1977 & 44 & \vline & 14-Feb-1983 & 2011 \\
27-Sep-1977 & 45 & \vline & 15-Feb-1983 & 2012 \\
28-Sep-1977 & 46 & \vline & 16-Feb-1983 & 2013 \\
29-Sep-1977 & 47 & \vline & 17-Feb-1983 & 2014 \\
30-Sep-1977 & 48 & \vline & 18-Feb-1983 & 2015 \\
4-Oct-1977 & 52 & \vline & 19-Feb-1983 & 2016 \\
30-Oct-1977 & 78 & \vline & 20-Feb-1983 & 2017 \\
10-Apr-1978 & 240 & \vline & 21-Feb-1983 & 2018 \\
11-Apr-1978 & 241 & \vline & 22-Feb-1983 & 2019 \\
17-Apr-1978 & 247 & \vline & 23-Feb-1983 & 2020 \\
24-Apr-1978 & 254 & \vline & 24-Feb-1983 & 2021 \\
25-Apr-1978 & 255 & \vline & 25-Feb-1983 & 2022 \\
26-Apr-1978 & 256 & \vline & 26-Feb-1983 & 2023 \\
27-Apr-1978 & 257 & \vline & 27-Feb-1983 & 2024 \\
28-Apr-1978 & 258 & \vline & 28-Feb-1983 & 2025 \\
1-May-1978 & 261 & \vline & 1-Mar-1983 & 2026 \\
2-May-1978 & 261 & \vline & 7-Mar-1983 & 2032 \\
28-Aug-1978 & 380 & \vline & 8-Mar-1983 & 2033 \\
29-Aug-1978 & 381 & \vline & 9-Mar-1983 & 2034 \\
30-Aug-1978 & 382 & \vline & 12-Mar-1983 & 2037 \\
31-Aug-1978 & 383 & \vline & 13-Mar-1983 & 2038 \\
1-Sep-1978 & 384 & \vline & 14-Mar-1983 & 2039 \\
2-Sep-1978 & 385 & \vline & 20-Mar-1983 & 2045 \\
3-Sep-1978 & 386 & \vline & 21-Mar-1983 & 2046 \\
4-Sep-1978 & 387 & \vline & 23-Mar-1983 & 2048 \\
2-Jan-1983 & 1968 & \vline & 26-Mar-1983 & 2051 \\
3-Jan-1983 & 1969 & \vline & 28-Mar-1983 & 2053 \\
4-Jan-1983 & 1970 & \vline & 29-Mar-1983 & 2054 \\
7-Jan-1983 & 1973 & \vline & 26-Apr-1983 & 2082 \\
13-Jan-1983 & 1979 & \vline & 9-May-1983 & 2095 \\
15-Jan-1983 & 1981 & \vline & 6-Dec-1984 & 2672 \\ [1ex]
\hline\hline 
\end{tabular}
\label{tab:tab1} 
\end{table*}

\bsp
\label{lastpage}
\end{document}